# INVESTIGATION OF 2-DIMENSIONAL FLUID FLOW USING FINITE DIFFERENCE FLOW METHOD OF NAVIER STOKES EQUATION

Dr. Mohit Kumar Srivastava^, Dr. Love Trivedi†, Rakshit Kaushik*

^Shri Vishwakarma Skill University
†* Department of Physics, JPIS, Jaipur, Rajasthan, India

**Abstract**

The presented research paper illustrates the development of a new methodology to solve 2-dimensional (2D) Navier-Stoke equations, which Pukhnachev proposed through introducing unknown functions in the stream and pressure functions of fluid flow. The proposed novel method is distinguishable from the common vorticity-stream given in Navier-Stokes expression because it has a stream function that corresponds to the unknown function in the elliptic expression. The equation represents a couple scheme in algorithmic considerations because it enables two situations to be solved using one function of the subject stream without putting new conditions on the innovative function. Here, the concept of numerical algorithm is applied in a flow under cavity to represent a benchmark task to be solved. The benchmark task gives enough representation of the subject flow.

**Introduction**

The Navier-Stoke equations can be solved using various numerical schemes when dealing with incompressible viscous flows. Some the numerous schemes are those that use variables commonly known as primitive or velocity-pressure and stream function. Others in the same category uses vorticity-steam and other formulations. Usually, finding numerical solutions when using the primitive variable poses a great challenge because the evolution expression will be missing in most cases, which can help solve the variable related to pressure. Therefore, the arising problem with using approach of primitive variables can be avoided by using vorticity-velocity and stream function vorticity formulations of the Navier-Stoke equation. However, the right values for vorticity boundary are not easy to determine. This study uses a new form of Navier-Stokes equation.

Pukhnachev proposed a new form of Navier-Stoke equation to solve 2D and axisymmetric flows. The novel Navier-Stoke expression in the 2D stream considering the unknown function has one expression for transportation that represent an elliptic mathematical expression and the functionality of the stream for the unknown. The stream function corresponds to both the stream and vorticity functions. However, the coupling function has a different physical meaning from the vorticity. A finite difference scheme is structed for the Navier-Stoke equation in the new methodology development. The algorithm considers the mathematical expression as a coupled organization and allows satisfaction of double situation for the function of the stream without putting condition on the undetermined function. It is possible to extend the proposed scheme to solve axisymmetric Navier-Stoke equations. The performance of this proposed method can be determined using a popular benchmark problem. One case that is applicable in the test is the simulation of a 2D lid-driven cavity flow under Reynolds number Re≥ 1000, under strict steady and laminar motion [3]. So far, research on numerical has revealed the properties of this cavity flow. Additionally, the flow of viscous stream in a forced cavity is one of the common test concepts in determining and verifying numerical problems.[3]

In the paper, several sections are included to illustrate various parts. The first part deals with the formulation of the proposed novel Navier-Stokes expression with no-slip boundary limits.

After that, a brief account of a task applied in the trial case is given with arithmetical algorithm. The next section gives the authentication findings of finite-different design with detailed comparison to tentative and numerical statistics.

**Formulation of Novel Navier-Stokes Equation**

It is essential to illustrate the steps involved in the formulation of novel equation for Navier-Stokes regarding incompressible viscosity in 2D, which makes the paper comprehensive. The Navier-Stokes expression governs the incompressible viscosity of the flow in a Cartesian coordinate presentation *(x, y)*,

$$u_t + uu_x + vu_y = -\frac{1}{\rho}p_x + \nu(u_{xx} + u_{yy})$$ ...........Equation (1)

$$v_t + uv_x + vv_y = -\frac{1}{\rho}p_y + \nu(v_{xx} + v_{yy})$$ ...........Equation (2)

$$u_x + v_y = 0$$ ...........Equation (3)

In the above equations, the parts of velocity are represented by *u* and *v* in *x* and *y* routes in that order. p= pressure, ρ = fluid density, and ν = kinematic viscosity. Notably, the fluid is under potential external forces[4]. Considering the flow of stream under 2D, the compressibility limit $\nabla \cdot v = 0$ can be satisfied through expressing vector of speed regarding the purpose of stream ψ according to:

$$u = \frac{\partial \psi}{\partial y}, \quad v = -\frac{\partial \psi}{\partial x}$$ ...........Equation (4)

From here, a new design of the popular Navier-Stokes equation can be pegged to the following scrutiny. When equation (4) is substituted in equation (1), it gives:

$$\frac{\partial}{\partial y}(\psi_t - \psi_x\psi_y - \nu\Delta\psi) + \frac{\partial}{\partial x}\left(\frac{1}{\rho}p + \psi_y^2\right) = 0 \quad\quad\quad\quad\quad\quad\quad\quad\quad\quad\quad\quad\quad\quad \text{Equation (5)}$$

Where $\Delta \stackrel{\text{def}}{=} \frac{\partial^2}{\partial x^2} + \frac{\partial^2}{\partial y^2}$.

Therefore, a function $\phi$ or ($\Phi$) satisfies the equation.

$$\frac{1}{\rho}p = -\psi_y^2 + \Phi_y \quad\quad\quad\quad\quad\quad\quad\quad\quad\quad\quad\quad\quad\quad\quad\quad\quad\quad\quad\quad\quad\quad\quad\quad \text{Equation (6)}$$

and

$$\psi_t - \psi_x\psi_y + \Phi_x = \nu\Delta\psi \quad\quad\quad\quad\quad\quad\quad\quad\quad\quad\quad\quad\quad\quad\quad\quad\quad\quad\quad\quad\quad\quad \text{Equation (7)}$$

When equations (6) and (7) are differentiated with respect to $y$ and $x$, and the result substituted into equation (2), where $u$ and $v$ are expressed in terms of $\psi$, it yields

$$\Delta\Phi = 2\psi_y\Delta\psi \quad\quad\quad\quad\quad\quad\quad\quad\quad\quad\quad\quad\quad\quad\quad\quad\quad\quad\quad\quad\quad\quad\quad\quad\quad\quad \text{Equation (8)}$$

Here, the case that correspond the no-slip state is considered as the limit of the flow sphere. Considering ψ only as a function of the stream, the conditions of the limits are

$$\psi = 0, \quad \frac{\partial\psi}{\partial n} = b(x,y) \quad\quad\quad\quad\quad\quad\quad\quad\quad\quad\quad\quad\quad\quad\quad\quad\quad\quad\quad\quad\quad \text{Equation (9)}$$

Where $\frac{\partial\psi}{\partial n}$ represent a vector with normal derivative towards the direction of the boundary. The formulation of the task can be completed through identifying the original forms

$$\psi = \psi_0(x,y), \quad t = 0 \quad\quad\quad\quad\quad\quad\quad\quad\quad\quad\quad\quad\quad\quad\quad\quad\quad\quad\quad\quad\quad\quad\quad \text{Equation (10)}$$

The key objective here is to develop and confirm a finite-difference design for solving the system (7) to (10) [1][2].

**Numerical Technique**

The lid-driven cavity flow is the standard problem for benchmarking verifying 2D Navier-Stokes equations shown below (*Figure 1*). The upper wall of the cavity pushes the fluid and results in a steady motion towards the right from left [4]. Let $L$ represent the characteristic scale related to the moving boundary.

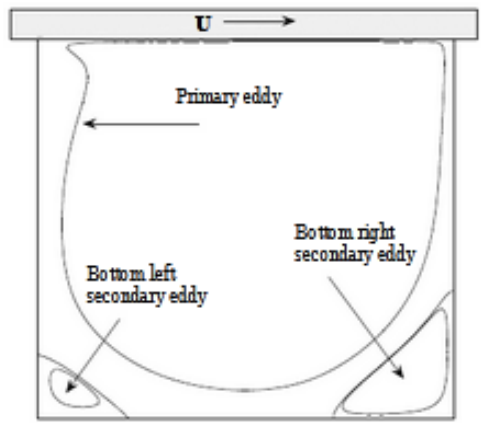

*Figure 1: Configuration of lid-driven square cavity flow*

The non-dimensional parameter of the problem is

$$Re = \frac{LU}{\nu}$$ ..........................................................................................................Equation (11)

The system of equations (1) to (3) is rendered dimensionless as follows

$$x = \frac{x^*}{L}, \quad y = \frac{y^*}{L}, \quad t = \frac{t^* \nu}{L^2}, \quad u = \frac{u^*}{U}, \quad v = \frac{v^*}{U}$$ ..........................................................Equation (12)

The domain is covered $Q = \{0 \leq x \leq 1, \ 0 \leq y \leq 1\}$ using a uniform grid

$Q_h = \{(x_i, y_j) | x_i = (i-1.5)h_x, \ y_j = (j-1.5)h_y, i = 1, \ldots, N_x, \ j = 1, \ldots, N_y\}$ with spacing

$h_x = \frac{1}{N_x - 2}$, $h_y = \frac{1}{N_y - 2}$, in the directions of $x$ and the $y$ axes accordingly. The given grid allows the use of median difference in approximating the conditions of the boundary with second-order on double-point patterns.

The suggested algorithm has one essential element that is considered as a pair of systems given in mathematic expressions (7) and (8) for $\psi$ and $\Phi$. The new formulation used here is based on the concept of the two border situations for $\psi$ as real situations for the $\psi$ - $\Phi$ system. Notably, $\psi$ and $\Phi$ are assessed using permanent procedure. The approximation for second-order central difference is used for the operators in equation (7) and (8). The difference equation for the system becomes

$$\frac{v_{i,j}^{n+1} - v_{i,j}^n}{\tau} - Re\frac{(v_{i+1,j}^n - v_{i-1,j}^n)(v_{i,j+1}^{n+1} - v_{i,j-1}^{n+1}) + (v_{i+1,j}^{n+1} - v_{i-1,j}^{n+1})(v_{i,j+1}^n - v_{i,j-1}^n)}{8 h_x h_y} +$$

$$Re\frac{(\Phi_{i+1,j}^{n+1} - \Phi_{i-1,j}^{n+1})}{2 h_x} - \frac{1}{2}(\Delta v_{i,j}^{n+1} + \Delta v_{i,j}^n)$$ ...........Equation (13)

$$\Delta \Phi_{i,j}^{n+1} = \frac{1}{2 h_y}\left[(v_{i,j+1}^n - v_{i,j-1}^n)\Delta v_{i,j}^{n+1} + (v_{i+1,j}^{n+1} - v_{i-1,j}^{n+1})\Delta v_{i,j}^n\right],$$

$i = 2, \ldots, N_x - 1$, $j = 1, \ldots, N_y - 1$. ..............Equation (14)

The conditions for the boundary are written as given below:

$$\frac{v_{2,j}^{n+1} + v_{1,j}^{n+1}}{2} = 0, \quad \frac{v_{2,j}^{n+1} - v_{1,j}^{n+1}}{h_x} = 0,$$

$$\frac{v_{N_x,j}^{n+1} + v_{N_x-1,j}^{n+1}}{2} = 0, \quad \frac{v_{N_x,j}^{n+1} - v_{N_x-1,j}^{n+1}}{h_x} = 0, \quad j = 2, \ldots, N_y - 1,$$

$$\frac{v_{i,2}^{n+1} + v_{i,1}^{n+1}}{2} = 0, \quad \frac{v_{i,2}^{n+1} - v_{i,1}^{n+1}}{h_y} = 0,$$

$$\frac{v_{i,N_y}^{n+1} + v_{i,N_y-1}^{n+1}}{2} = 0, \quad \frac{v_{i,N_y}^{n+1} - v_{i,N_y-1}^{n+1}}{h_y} = 1, \quad i = 2, \ldots, N_x - 1.$$

...........Equation (15)

It is possible to add (13) to (15) and obtain one linear scheme that is banded with matrix. The matrix enables an introduction of a new scheme of indices.

$$k_{(i,j)} = 2(j-1)N_x + 2i - 1, \quad i = 1, \ldots, N_x,$$
$$m_{(i,j)} = 2(j-1)N_x + 2i = k_{(i,j)} + 1, \quad j = 1, \ldots, N_y.$$

Each of the nodules of grid $Q_h$, especially $(i, j)$ is associates with indices including $k(i,j)$ and even $m(i,j)$. Note that the index $k(i,j)$ represents an odd number while $m(i,j)$ represents an even number. Therefore,

$$k_{(i+1,j)} - k_{(i,j)} + 2, \quad k_{(i,j+1)} - k_{(i,j)} + 2N_x,$$
$$k_{(i-1,j)} - k_{(i,j)} - 2, \quad k_{(i,j-1)} - k_{(i,j)} - 2N_x,$$
..........Equation (16a)

$$m_{(i+1,j)} - m_{(i,j)} + 2, \quad m_{(i,j+1)} - m_{(i,j)} + 2N_x,$$
$$m_{(i-1,j)} - m_{(i,j)} - 2, \quad m_{(i,j-1)} - m_{(i,j)} - 2N_x.$$
..........Equation (16b)

A new grid function $\sigma - \{\sigma_k, k - 1, ..., 2N_x N_y\}$ can be introduced as defined in the composite grid. Taking the components of the grid function to be $\sigma_k$ with even indices representing $\psi_{i,j}$ and components with odd indices $\sigma_m(-\sigma_{k+1})$ to represent $\Phi_{i,j}$ can be substituted instead of $\psi_{i,j}$ and $\sigma_m$ substituted instead of $\Phi_{i,j}$ in equation (13) and (14), which allows recasting of algebraic system as follows

$$\frac{\sigma_k^{n+1} - \sigma_k^n}{\tau} + \frac{Re}{8h_x h_y}\left[(\sigma_{k+2}^n - \sigma_{k-2}^n)(\sigma_{k+2N_x}^{n+1} - \sigma_{k-2N_x}^{n+1}) + (\sigma_{k+2}^{n+1} - \sigma_{k-2}^{n+1})\right.$$
$$\left.(\sigma_{k+2N_x}^n - \sigma_{k-2N_x}^n)\right] - \frac{Re}{2h_x}(\sigma_{k+1}^{n+1} - \sigma_{k-3}^{n+1}) - \frac{1}{2}(\Delta\sigma_k^{n+1} + \Delta\sigma_k^n)$$
..........Equation (17a)

$$\Delta\sigma_m^{n+1} - \frac{(\sigma_{m+2N_x-1}^n - \sigma_{m-2N_x-1}^n)}{2h_y}\Delta\sigma_{m-1}^{n+1} + \frac{(\sigma_{m+2N_x-1}^{n+1} - \sigma_{m-2N_x-1}^{n+1})}{2h_y}\Delta\sigma_{m-1}^n$$
..........Equation (17b)

Where $\Delta\sigma_k - \frac{(\sigma_{k+2} - 2\sigma_k + \sigma_{k-2})}{h_x^2} + \frac{(\sigma_{k+2N_x} - 2\sigma_k + \sigma_{k-2N_x})}{h_y^2}.$

A simple implementation of the algorithm results in a problem with a single numerical matrix. The resultant problem can be standardized in different ways. Notably, addition of a small element to the boundary gives the best results [5]. Using this idea, equation (15) can be rearranged for function $\sigma$

$$\frac{\sigma_k^{n+1} + \sigma_{k+2}^{n+1}}{2} - 0, \quad \frac{\sigma_{m+1}^{n+1} - \sigma_{m-1}^{n+1}}{h_x} - \varepsilon\sigma_m^{n+1}, \quad i - 1, j - 1, \ldots, N_y,$$
..........Equation (18a)

$$\frac{\sigma_k^{n+1} + \sigma_{k-2}^{n+1}}{2} - 0, \quad \frac{\sigma_{m-1}^{n+1} - \sigma_{m-3}^{n+1}}{h_x} - \varepsilon\sigma_m^{n+1}, \quad i - N_x, j - 1, \ldots, N_y,$$
..........Equation (18b)

$$\frac{\sigma_k^{n+1} + \sigma_{k+2N_x}^{n+1}}{2} - 0, \quad \frac{\sigma_{m+2N_x+1}^{n+1} - \sigma_{m-1}^{n+1}}{h_y} - 0, \quad j - 1, i - 1, \ldots, N_x,$$
..........Equation (18c)

$$\frac{\sigma_k^{n+1} + \sigma_{k-2N_x}^{n+1}}{2} = 0, \quad \frac{\sigma_{m-1}^{n+1} - \sigma_{m-2N_x-1}^{n+1}}{h_y} = 1, \quad j = N_y, \; i = 1, \ldots, N_x,$$

...............................Equation (18d)

Where ε is a small number. For steady flow. The algorithm is considered as an interactive step and the interaction is terminated at *n-N* satisfying the following criteria.

$$\frac{\max_{i,j} |\sigma_{i,j}^{N+1} - \sigma_{i,j}^{N}|}{\max_{i,j} |\sigma_{i,j}^{N+1}|} \leqslant 10^{-8}$$

Notably, the coupling of the system results in the formation of a linear system $\psi - \Phi$, which can be rearranged based on multi-diagonal arrangement for the function of grid composite $\sigma$

$$B_{l-2N_x-1}\sigma_{l-2N_x-1}^{n+1} + B_{l-2N_x}\sigma_{l-2N_x}^{n+1} + B_{l-3}\sigma_{l-3}^{n+1} + B_{l-2}\sigma_{l-2}^{n+1}$$
$$+ B_{l-1}\sigma_{l-1}^{n+1} + B_l\sigma_l^{n+1} + B_{l+1}\sigma_{l+1}^{n+1} + B_{l+2}\sigma_{l+2}^{n+1}$$
$$+ B_{l+2N_x-1}\sigma_{l+2N_x-1}^{n+1} + B_{l+2N_x}\sigma_{l+2N_x}^{n+1} + B_{l+2N_x+1}\sigma_{l+2N_x+1}^{n+1} = F_l,$$

...............................Equation (19)

Where $l = 1, \ldots, 2N_xN_y$. Equation 19 gives a linear system with matrix that is bound with $2N_x + 1$ on both upper side and lower side. The problem given in equation 19 can be solved using routings of LAPACK.

**Results and Discussion**

A confirmation test involve flow in 2D cavity at Reynolds number of up to 1000, where they give steady and laminar flow. Calculations were done for the lid-driven cavity task on grid from 32 * 32 to 102 * 102. ε, which represents a small number, has a sensible impact on the findings of the test, which were evaluated using the numerical analysis. The analysis showed that

$\varepsilon \in [10^{-10}, 10^{-4}]$. The results prove that the solution was compatible with the results of the test cases. The validation of the subject scheme was done through comparing values of external points and those of locations in space functions of the target stream. The target stream was having [2,3,4,5, and 6]. Consider table 1. The top row shows values form simulation while the bottom row shows values from other sources. For the primary vortex, the results showed agreement within 5% as compared to values from other sources. For the secondary vortex, the 52 * 52 grid, portrayed results agree within tolerable 5% [5]. However, the values were different as compared to (Sellountos et al., 2019)[6]. The data from 102 * 102 grid revealed perfect match with those data from [2,3,4,6]. The structures for geometry of the flow are shown in figures 2 and 3 using the 102 * 102 grid. Notably, the values of velocity (*u*) along x⁻ direction, velocity (*v*) along y⁻ axis, and *w* were calculated from the ψ after the convergence of the interaction. The values of *u, v,* and *w* in the domain were approximated using central difference and the same values on the boundary were approximated using difference of one side first-order. Figure 4 reveals a comparison between centerline *u⁻* and *v⁻* profiles for velocities.

| Ref. | Grid | Primary eddy | | Bottom right second. eddy | |
|---|---|---|---|---|---|
|  |  | $\psi_{min}$ | $(x_{min}, y_{min})$ | $\psi_{max}$ | $(x_{max}, y_{max})$ |
|  |  | Re = 100 |  |  |  |
| Present | 32 × 32 | −0.103615 | (0.600, 0.733) | 3.34039 × 10⁻⁶ | (0.967, 0.067) |
|  | 52 × 52 | −0.103569 | (0.620, 0.740) | 9.42029 × 10⁻⁶ | (0.940, 0.060) |
|  | 102 × 102 | −0.103510 | (0.620, 0.740) | 1.18920 × 10⁻⁵ | (0.940, 0.060) |
| [2] | 48 | −0.10008 | − | − | − |
| [4] | 162 × 162 | −0.10397 | (0.6198, 0.7369) | − | − |
| [6] | 129 × 129 | −0.103 | (0.5844, 0.7400) | 1.25 × 10⁻⁵ | (0.9401, 0.0599) |
|  |  | Re = 1000 |  |  |  |
| Present | 52 × 52 | −0.119825 | (0.540, 0.560) | 1.56192 × 10⁻³ | (0.860, 0.100) |
|  | 102 × 102 | −0.119280 | (0.530, 0.560) | 1.68817 × 10⁻³ | (0.870, 0.110) |
| [2] | 160 | −0.118937 | (0.5308, 0.5652) | 1.72972 × 10⁻³ | (0.8640, 0.1118) |
| [3] | 1024 × 1024 | −0.11892 | (0.5312, 0.5654) | 1.7292 × 10⁻³ | (0.8643, 0.1123) |
| [4] | 80 × 80 | −0.118710 | (0.5346, 0.5645) | − | − |
| [5] | 512 × 512 | −0.116269 | (0.5316, 0.5660) | 1.640 × 10⁻³ | (0.8651, 0.1118) |
| [6] | 128 | −0.117929 | (0.5313, 0.5625) | 1.751 × 10⁻³ | (0.8594, 0.1094) |

Table 1: Strength and location of vortices at Re – 100 to 1000

It is notable that understanding the behavior of function $\Phi$ for various Re. Figure 5 illustrates the contour of the function $\Phi$ for several Re and the set of figures given from 52 * 52 grid with parameters $\varepsilon - 10^{-6}$.

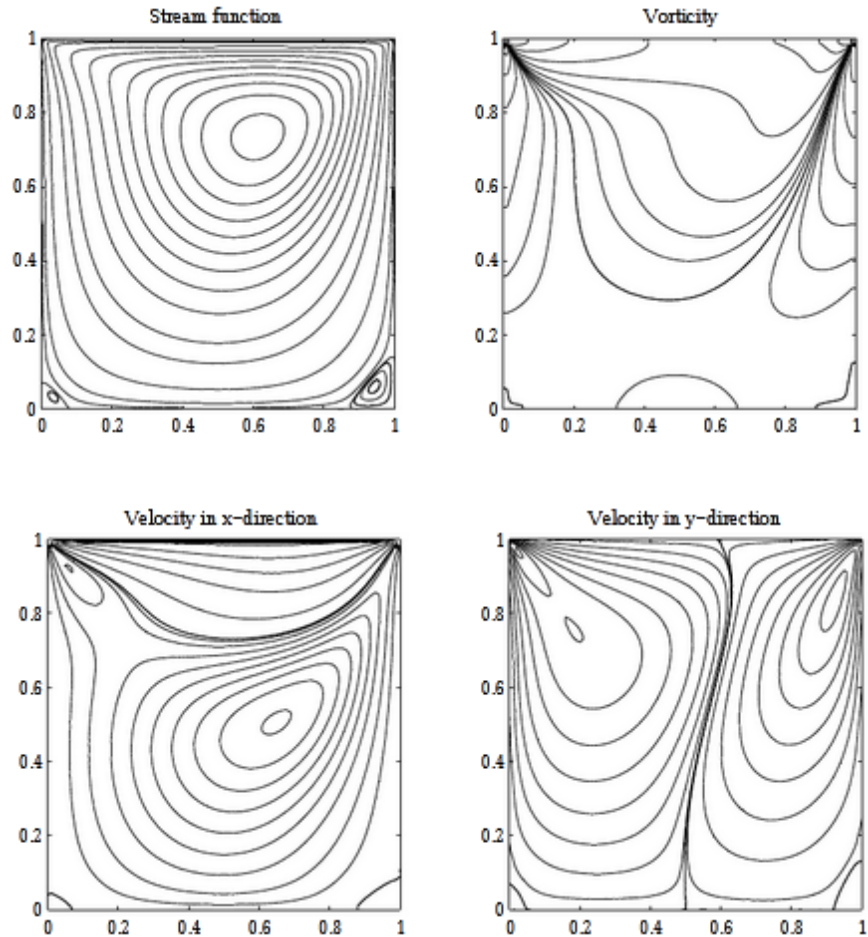

Figure 2: Functions of stream ψ, vorticity w, and velocity components *u* and *v* contours for the lid-driven flow at Re-100.

**Conclusion**

Remarkably, the development of finite difference sachem was successful and it confirmed the validity of the proposed novel formulation using the 2D Navier-Stokes equation. The development revealed that the stream function was reliable enough to be used in evaluating the equation, which is a distinguishing property of the novel development in the application of Navier-Stokes equation. It is notable that the developed algorithm and its counterpart scheme considered the subject equation as a coupled system, and it allows satisfaction of two conditions when dealing with functions when dealing with function streams without putting new conditions on the derived function. The formulation portrayed high accuracy and reliable efficiency for the benchmark problem in lid-driven cavity flow.